\documentclass[twocolumn,showpacs,preprintnumbers,amsmath,amssymb,prl]{revtex4}

\usepackage{dcolumn}
\usepackage{bm}
\usepackage{hyperref} 
\usepackage{amsmath}
\usepackage{amsfonts}     
\usepackage{graphics}  
\usepackage{psfrag} 
\usepackage{graphicx}    
\usepackage{epsfig}    
\usepackage{rotating}  
 \usepackage[latin1]{inputenc}
 \usepackage{color}
 \usepackage{rotating}

\begin{document}


\title{Zero-one survival behavior of cyclically competing species}

\author{Maximilian Berr$^{1,2}$}
\author{Tobias Reichenbach$^{1}$}
\author{Martin Schottenloher$^{2}$}
\author{Erwin Frey$^{1}$}

\affiliation{$^{1}$Arnold Sommerfeld Center for Theoretical Physics (ASC) and Center for NanoScience (CeNS), LMU M\"unchen, Theresienstra{\ss}e 37, 80333 M\"unchen, Germany\\
$^{2}$Mathematisches Institut der LMU M\"unchen, 
Theresienstra{\ss}e 39, 80333 M\"unchen, Germany
}


\date{\today}
\begin{abstract}
Coexistence of competing species is, due to unavoidable fluctuations, always transient. In this Letter, we investigate the ultimate survival probabilities characterizing  different species in cyclic competition.  We show that they often obey a surprisingly simple, though non-trivial behavior. Within a model where coexistence is neutrally stable, we demonstrate a robust zero-one law: When the interactions between the three species are (generically) asymmetric, the `weakest' species survives at a probability that tends to one for large population sizes, while the other two are guaranteed to extinct. We rationalize our findings from stochastic simulations by an analytic approach.
\end{abstract}

\maketitle

Ecological systems are composed of a large number of different interacting species~\cite{May}. Competition between them basically affects the probability of individuals' reproduction as well as  death. However, such birth and death processes also possess a considerable degree of stochasticity, which induces fluctuations that ultimately result in species extinction~\cite{Gause,Hubbell}. 
In this respect, further understanding of the conditions and mechanisms that enable the huge observed biodiversity is subject of a large body of work in contemporary theoretical ecology and biological physics. 
It involves the challenging problems of characterizing  out-of-equilibrium steady states in the presence of intrinsic fluctuations~\cite{traulsen-2005-95}, nonlinearities~\cite{Hofbauer} as well as nontrivial interaction networks~\cite{szabo-2007-446}.  
Recent experiments on colicinogenic microbes have proven the importance of cyclic, `rock-paper-scissors' like,  competition~\cite{kerr-2002-418}.  Such cyclic dynamics also governs, \emph{e.g.},  certain lizard populations~\cite{sinervo-1996-340} and coral reef invertebrates~\cite{jackson-1975-72}. Theoretical studies have mainly focussed on identifying conditions under which cyclic competition leads to maintained diversity, employing, \emph{e.g}., a time-scale framework to distinguish stable from unstable coexistence~\cite{reichenbach-2007-448}. A supporting role of self-forming spatial patterns has been underlined generally~\cite{hassell-1994-370,durrett-1998-53,reichenbach-2007-448}, although in certain situations it may not be necessary~\cite{claussen-2008-100} or occasionally even harmful~\cite{reichenbach-2008-101}. 

In contrast, little is known about the fingerprints of extinction. E.g., in the  \emph{E.coli} experiments~\cite{kerr-2002-418}, when the population is well-mixed,  a strain which is resistant to the poison without producing toxin itself remains as the only survivor after a short transient. Is this behavior robust? If so, why does one species reliably outcompete the other two although all three species together display a cyclic dynamics, where each outcompetes another but is itself beaten by the remaining one? What is the influence of unavoidable fluctuations?

In this Letter, we approach these ecologically important and physically insightful questions by investigating cyclic competition of three interacting species, referred to as $A,B$, and $C$.  Aiming at a  broad and general applicability of our model and results, we consider the following simplified, paradigmatic interactions:
\begin{eqnarray}
&A + B\stackrel{k_A}{\longrightarrow}  A + A\,,   \cr
&B + C\stackrel{k_B}{\longrightarrow}  B + B\,, \cr
&C + A\stackrel{k_C}{\longrightarrow}  C + C\,. \label{RS}
\end{eqnarray}
Hereby, $A$ outperforms $B$ at a rate $k_A$, while $B$ beats $C$ which outcompetes $A$ in turn, at rates $k_B$ resp. $k_C$. Recently, it has been shown that a population of $N$ such interacting individuals eventually ends up in one of the (absorbing) states  where only one species survives~\cite{ifti-2003-10,reichenbach-2006-74}. The mean time $T$ for extinction is  proportional to the system size $N$, $T\sim N$, indicating that extinction is solely driven by fluctuations~\cite{ifti-2003-10,reichenbach-2006-74}. Therefore, which one of the species survives is subject to a random process. If the interaction rates are equal, \emph{i.e.}, the three species are symmetric, all have an equal chance of surviving.

\begin{figure}[t]   
\begin{center}    
\includegraphics[width=6cm]{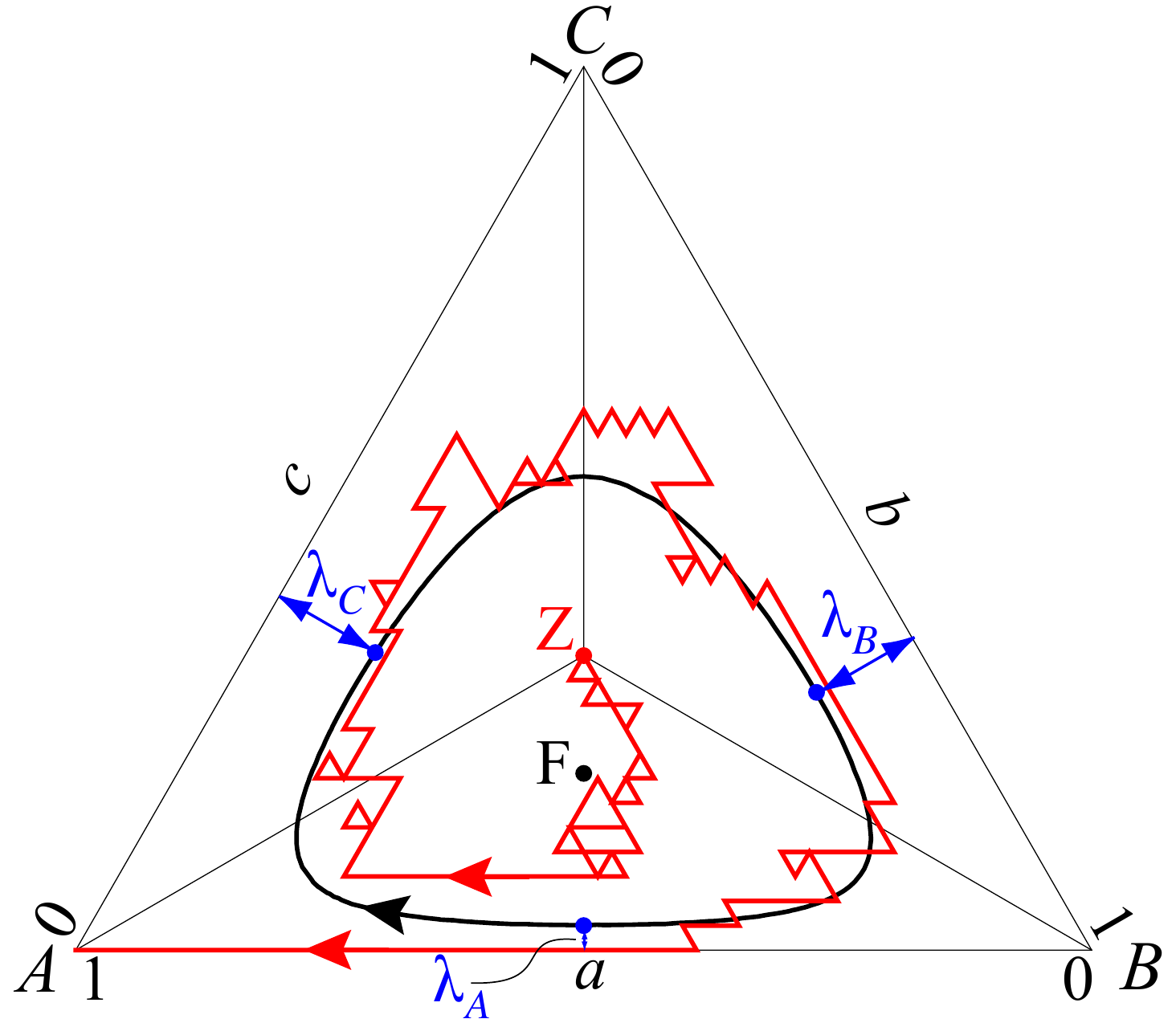}
\caption{The phase space $S_3$. We show the reactive fixed point F, the center Z, as well as a stochastic trajectory (red/light gray). It eventually deviates from the `outermost'  deterministic orbit (black) and reaches the absorbing boundary. For the distances $\lambda_A,~\lambda_B$ and $\lambda_C$ (blue/dark gray) see text.
Parameters are $(k_A, k_B, k_C) = (0.2,0.4, 0.4)$ and $N=36$.}
\label{SSZW}
\end{center}                
\end{figure}

\begin{figure}[t]   
\begin{center}    
\includegraphics[width=8.4cm]{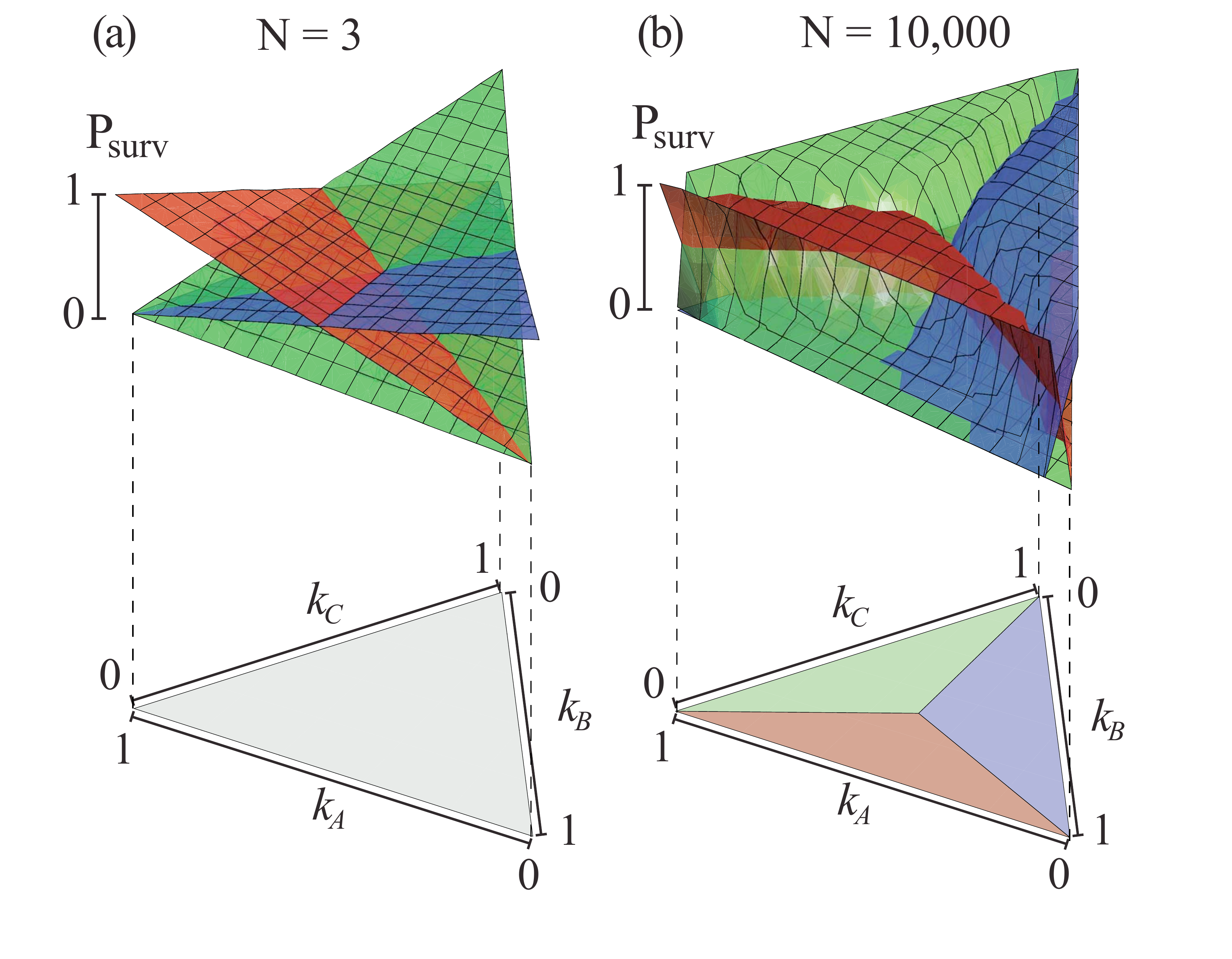}
\caption{Survival probabilities for $A$ (blue/dark gray), $B$ (green/medium gray) and $C$ (red/light gray), obtained from stochastic simulations as averages over $10,000$ samples. (a), A small system, $N=3$, leads to a `law of stay-out' and linear dependences of $\text{P}_\text{surv}$  on the reaction rates. (b), Large populations, here for $N=10,000$, are governed by a `law of the weakest' and obey, in the limit $N\to\infty$, a zero-one behavior.
}
\label{SIMDATA}
\end{center}                
\end{figure}

Here, we investigate the generic case where the competing species do not obey a symmetry. To gain intuition for the system's behavior, we discuss predictions by the rate equations (RE) first. The RE describe the deterministic time-evolution of the densities $a,b$ and $c$ of the three species, as may arise when fluctuations are negligible, \emph{e.g.}, when the population size $N$ is large. For the reactions~(\ref{RS}), the RE are given by
\begin{eqnarray}
&\dot{a}=a(k_Ab-k_Cc)\,, \cr
&\dot{b}=b(k_Bc-k_Aa)\,, \cr 
&\dot{c}=c(k_Ca-k_Bb)\,.  \label{DE} 
\end{eqnarray}
They  conserve the number $N$ of interacting individuals: the densities fullfill the relation $a+b+c=1$, spanning the simplex $S_3$ as the phase space (see Fig.~\ref{SSZW}).
Its corners represent three absorbing fixed points, where only one species remains. In addition, Eqs.~(\ref{DE}) possess a reactive fixed point F, located at $(a^*,b^*,c^*)=(k_B,k_C,k_A)/(k_A+k_B+k_C)$ which corresponds to species coexistence. In the following, we  use the time-scale normalization condition $k_B+k_C+k_A=1$. Therewith, the parameter space, spanned by the rates $k_A,k_B,k_C$,  also adopts the form of the  simplex $S_3$, see the lower parts of Fig.~\ref{SIMDATA}.

The RE~(\ref{DE}) predict neutral stability of species coexistence, as they obey the following constant of motion:
\begin{equation}
R\equiv a^{k_B} b^{k_C}c^{k_A}\,, 
\label{COM}
\end{equation}
which does not change in the course of the deterministic time-evolution. Similar to the energy in classical mechanics, $R$ singles out closed orbits surrounding the coexistence fixed point F, see  Fig.~\ref{SSZW}. These orbits, as well as F, are neutrally stable to fluctuations; stochastic trajectories follow the cyclic behavior of the deterministic orbits to a certain degree, while at the same time performing a random walk between them. Eventually they reach the boundary of the phase space and are then driven to one of the absorbing fixed points (c.f. Fig.~\ref{SSZW}).

We have performed extensive computer simulations to determine the influence of the reaction rates $k_A,k_B,k_C$ as well as the system size $N$  on the probabilities $\text{P}_\text{surv}$ for each species to survive.  To this end we have evolved the system, with initial coexistence, until extinction of two species occurred; the average outcome over many such runs defined the survival probabilities. Typically, the system has initially been in a state corresponding to the center of the simplex. However, altering this starting point is not relevant for the results, as we show below.

What is the influence of the population size $N$ on the survival behavior? To answer this question, firstly, we consider
the smallest population where all three species can `coexist', namely $N=3$. In this case, $\text{P}_\text{surv}$ depend linearly on the reaction rates, see Fig.~\ref{SIMDATA} (a). For such a small system, the master equation describing the stochastic processes~(\ref{RS}) can be solved exactly. Only the state where one individuals of each species is present corresponds to coexistence, the other states lie on the absorbing boundary. A single process  therefore immediately leads to extinction and determines which species survives. The resulting survival probability for species $A$ reads  $\text{P}^A_\text{surv}=k_B$, the others follow analogously. A `law of stay-out' arises: The species that is least frequently engaged in interactions (for species $A$, interactions occur at rates $k_A$ and $k_C$) has the highest chance to survive. In contrast to what emerges for large populations, see below, this law is not strict. If $k_B$ denotes the largest of the interaction rates, species $A$ is not guaranteed to survive, but possesses the highest probability.

Large populations, about $N>100$, display a contrasting `law of the weakest' which determines the surviving species~\cite{frean-2001-268}. Namely, for reaction rates fullfilling $k_A<k_B, k_C$, species $A$ has the highest probability of surviving, although $A$ may be considered as the `weakest' species: Its reproduction occurs at rate $k_A$ and is thus the slowest of the three competing species. This nontrivial law has previously been described, as a non-strict one, in Ref.~\cite{frean-2001-268}. Here we show that, surprisingly, this law becomes strict in the limit of large population sizes, $N\to\infty$ (see Fig.~\ref{SIMDATA} (b) for the situation $N=10,000$). In this limit, $\text{P}^A_\text{surv}\to 1$ in the region $k_A<k_B,k_C$, and $\text{P}^A_\text{surv}\to 0$ otherwise; a `zero-one' behavior arises. Three regions emerge in parameter space, depicted in the lower part of Fig.~\ref{SIMDATA} (b). In each of them one distinct  species is guaranteed to survive, while the others go extinct. 

The transition from the `law of stay-out' to the `law of the weakest' happens gradually upon increasing the system size. From our stochastic simulations,  small populations, around $N<20$, are governed by the `law of stay-out' (although strictly valid only for $N=3$). Intermediate scenarios emerge for medium populations, about $20<N<100$, while survival in large ones, $N>100$, is  predominantly determined by the `law of the weakest'.

\begin{figure}[t]   
\begin{center}    
\includegraphics[width=8cm]{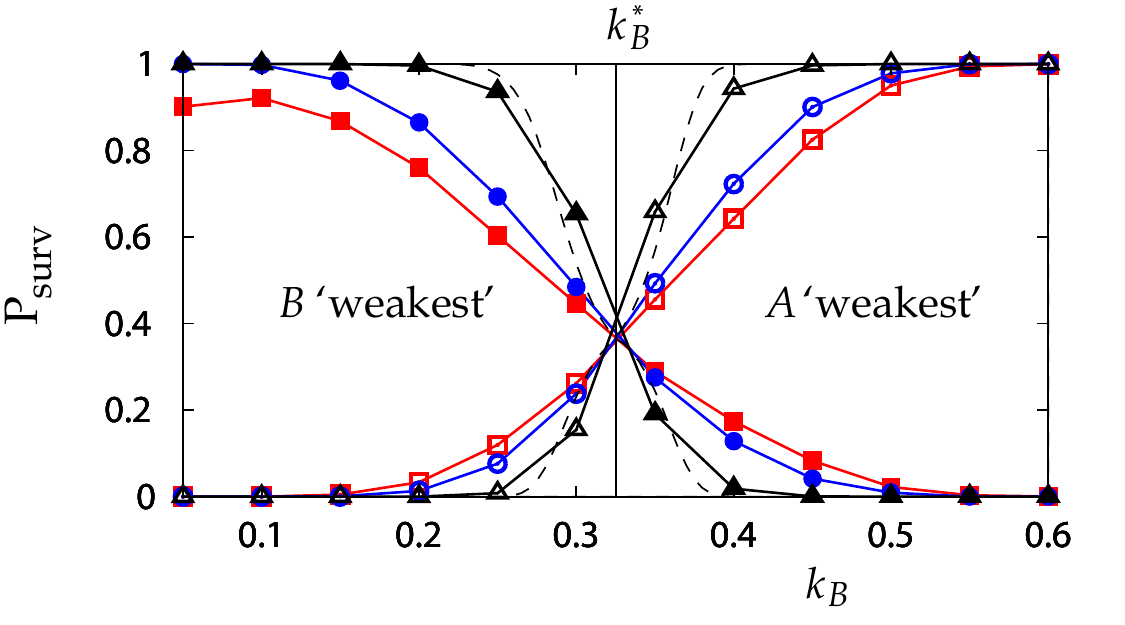}
\caption{The limit of large populations. We show the survival probabilities of $A$ (open symbols) and $B$ (filled) depending on $k_B$, for $k_C=0.35$ and different system sizes: $N=24$ (red squares),  $100$ (blue circles), and $10,000$ (black triangles). A sharpened transition emerges at $k_B^*=0.325$ where the $A$- and $B$-dominated regions meet. Our analytical prediction, Eq.~(\ref{surv_prob}) is shown for $p=0.7$ and $N=10,000$ (dashed line).}
\label{SV}
\end{center}                
\end{figure} 
We have confirmed that the `law of the weakest' becomes strict for $N\to\infty$ by computing the survival probability for different population sizes $N$. For illustration, let us focus on a one-dimensional section through the two-dimensional parameter space, connecting a region where $A$ is `weakest' to another one where $B$ is `weakest'. Results are shown in Fig.~\ref{SV}; they  demonstrate that the transition between $A$- and $B$-survival, at the `critical' parameter value where both regions meet, becomes sharp when increasing the population size, resulting in a  discontinuous transition.

Further evidence for a sharp transition in the survival probabilities, \emph{i.e.}, a zero-one behavior, comes from an analytical approach which we have developed. It  describes the survival probabilities for large systems, where only the last part of the stochastic time-evolution, shortly before extinction, is relevant. Indeed, in large systems, the stochastic trajectories exhibit only small fluctuations, and closely follow the deterministic orbits, performing many turns around the reactive fixed point F (see Fig.~\ref{SSZW}). Eventually, they deviate from an `outermost' deterministic orbit, shown in black in Fig.~\ref{SSZW}, and hit the absorbing boundary. Then, the system is driven to one of the absorbing states. Which one is reached depends on which edge of phase space the trajectory had reached before; each edge leads to one distinct uniform state. 

For asymmetric reaction rates, the fixed point is shifted from the center Z of the phase space towards one of the three edges, \emph{i.e.}, into one of the three domains shown in Fig.~\ref{SSZW}. All deterministic surrounding orbits are changed in the same way, squeezing in the direction of one edge.  Intuitively, the absorbing state which is reached from this edge has the highest probability of being hit, as the distance from the `outermost' deterministic orbit towards this edge is shortest. Indeed, this behavior has been validated by the above presented stochastic simulations.

Let us formalize the above considerations. We define the outermost deterministic orbit as the one orbit that is only a distance of $1/N$, \emph{i.e.}, one discrete, elementary step, apart from its closest edge. The distance of this outermost orbit to the edge that induces survival of species $A$ is termed $\lambda_A$; the distances $\lambda_B$ and $\lambda_C$ are defined analogously. Now, in the parameter region $k_A<k_B,k_C$, where $A$ has the highest survival probability,  the distance $\lambda_A$ is smallest, and therefore $\lambda_A=1/N$. The other two distances can be obtained via the conserved quantity $R$, Eq.~(\ref{COM}). For this purpose, in the following, we consider the (most interesting) situation where the differences between the reaction rates $k_A,k_B,k_C$ are small. The outermost orbit then runs through the point $c=\lambda_A=1/N$ and $a\approx b\approx 1/2$, yielding its constant of motion 
\begin{equation}
R^\text{o.O}=\frac{1}{N^{k_A}}\frac{1}{2^{k_B+k_C}}\,.
\end{equation}
If we perform the same calculation at the point where the outermost orbit lies closest to the edge leading to survival of species $B$ resp. $C$, we obtain
\begin{equation}
R^\text{o.O}=\lambda_B^{k_B}\frac{1}{2^{k_C+k_A}}~~\text{and}~~R^\text{o.O}=\lambda_C^{k_C}\frac{1}{2^{k_B+k_A}}\,,
\end{equation}
resp.. Equating these expressions yields 
\begin{equation}
\lambda_B=2^\frac{k_A-k_B}{k_B}\times N^{-\frac{k_A}{k_B}}~~\text{and}~~\lambda_C=2^\frac{k_A-k_C}{k_C}\times N^{-\frac{k_A}{k_C}}\,.
\label{lambdas}
\end{equation}

Most important for our purpose is the scaling of the distances $\lambda_B,\lambda_C$ in the population size $N$. Residing within the regime $k_A<k_B,k_C$, we notice from Eqs.~(\ref{lambdas}) that both $\lambda_A$ and $\lambda_B$ decrease slower than $1/N$.  Consequently, the number of discrete steps that separate the outermost orbit from these two edges, given by $N\lambda_B$ resp. $N\lambda_C$, tend, for large populations, to infinity. Note that the same does not apply to $\lambda_A$, which we keep fixed at $1/N$. Below, we show how this scaling  leads to the zero-one behavior of the survival probabilities.

The probability for deviating from the outermost orbit and performing an elementary step towards the absorbing boundary of the phase space is, for small differences in the reaction rates, approximately constant along the orbit, let us denote it by $p<1$. Now, the probability of leaving the outermost orbit and reaching the edge leading to survival of species $B$ is given by the probability of $N\lambda_B$ such subsequent elementary steps, and therefore reads $p^{N\lambda_B}$. Analogously, we obtain $p^{N\lambda_A}$ and $p^{N\lambda_C}$ as the probabilities for reaching the edges connected to $A$ and $C$, resp.. Consequently, the survival probabilities of the different species are given by the corresponding normalized probabilities:
\begin{equation}
\text{P}^A_\text{surv}=\frac{p^{N\lambda_A}}{p^{N\lambda_A}+p^{N\lambda_B}+p^{N\lambda_C}}\,,
\label{surv_prob}
\end{equation}
$P^B_\text{surv}$ and $P^C_\text{surv}$ follow analogously. The above found scaling, $N\lambda_B, N\lambda_C\to\infty$ upon $N\to\infty$, while $N\lambda_A=1$, imply that for large populations
$\text{P}^A_\text{surv}\to 1,~\text{P}^B_\text{surv}\to 0,~\text{P}^C_\text{surv}\to 0$, and therefore the zero-one behavior emerges. Note that these results have been derived from assuming  $k_A<k_B,k_C$; different species survive in the other regions of the parameter space. The overall sigmoidal form of Eq.~(\ref{surv_prob}) and the associated width agrees well with numerical findings for large systems, see Fig.~\ref{SV}. However, deviations do occur in the more detailed shape of the survival probability. We attribute them to the  approximations we made when deriving Eq.~(\ref{surv_prob}), namely that we have treated the escape step probability $p$  along the outermost orbit as well as the latter's distances to the different edges as constant.

Symmetries between species alter the survival probabilities. If all interaction rates are identical, all species clearly have the same chance of $1/3$ to survive; and if two rates coincide, the corresponding two species can both have chance $1/2$ of remaining. 

The above analysis suggests that the survival probabilities do not depend on the starting point, as long as the latter is not too close to the absorbing boundary. Indeed, extinction occurs due to deviations from the outermost orbit, any initial state therein will induce the same behavior. This expectation has been confirmed by  simulations.

The dependence of the extinction behavior solely on temporally late deviations from the outermost orbit is reminiscent of `tail-events' in probability theory~\cite{Gut} which induce the celebrated zero-one-law originally due to  Kolmogorov~\cite{Gut}. While this law cannot be directly applied to the present situation, mainly due to the finite number of steps in each of the trajectories discussed here, further investigations along these lines seem promising to deepen our understanding of zero-one behaviors.

We have derived the zero-one behavior of survival probabilities, accompanied by a strict `law of the weakest', for a cyclic population model that, deterministically, exhibits neutrally stable coexistence. However, the above analysis based on scaling arguments allows us to immediately generalize the obtained results to the case where coexistence is (deterministically) stable. Then, stochastic trajectories are attracted to the reactive fixed point, with rare large deviations. Again, extinction is determined by the behavior of trajectories close to the absorbing boundary, such that an analogous analysis as derived above holds. However, the latter is much harder to test numerically. Deterministically stable coexistence induces a mean extinction time which increases exponentially in the system size~\cite{cremer-2008-63}, such that computation of survival probabilities in large systems is hardly feasible.

We conclude by relating our results to the  \emph{E.coli} experiments~\cite{kerr-2002-418} mentioned in the beginning. Identifying $A$
with the sensitive, $B$ with the resistant, and $C$ with the colicinogenic strain, we uncover the relation $k_C \gg k_A>k_B$ [$C$ can
kill $A$ (fast) and reproduce, while $A$ and $B$ can only reproduce if a neighboring bacterium dies (slow). $k_A$ and $k_B$ are then proportional to
the reproduction rate differences of $A$ and $B$ resp. of $B$ and $C$. The measured data~\cite{kerr-2002-418} leads to $k_A>k_B$ ]. The resistant strain $B$ is thus `weakest' and,
according to the `law of the weakest', survives, in agreement with experimental observations~\cite{kerr-2002-418}.

Financial support of the German Excellence Initiative via the program ``Nanosystems
Initiative Munich" and the German Research Foundation via the SFB TR12 ``Symmetries and Universalities in Mesoscopic Systems''  is gratefully acknowledged. T. R. acknowledges funding by the Elite-Netzwerk Bayern.



\end{document}